\newcommand{\be}{\begin{equation}}
\newcommand{\ee}{\end{equation}}
\newcommand{\bea}{\begin{eqnarray}}
\newcommand{\eea}{\end{eqnarray}}
\newcommand{\sn}{{\rm sn}}
\newcommand{\ds}{{\rm ds}}
\newcommand{\cs}{{\rm cs}}
\newcommand{\ns}{{\rm ns}}
\newcommand{\dn}{{\rm dn}}
\newcommand{\cn}{{\rm cn}}
\newcommand{\sech}{{\rm sech}}

\documentclass[showpacs,preprintnumbers,amsmath,amssymb]{revtex4}
\usepackage{graphicx}
\usepackage{dcolumn}
\usepackage{bm}

\begin{document}
\vspace{.5in}
\begin{center}
{\LARGE{\bf Exact Moving and Stationary Solutions of a Generalized Discrete
Nonlinear Schr\"odinger Equation}}
\end{center}

\vspace{.3in}
\begin{center}
{\LARGE{\bf Avinash Khare}} \\
{Institute of Physics, Bhubaneswar, Orissa 751005, India}
\end{center}

\begin{center}
{\LARGE{\bf Sergey V. Dmitriev}} \\
{General Physics Department, Altai State Technical University,
Barnaul 656038, Russia}
\end{center}

\begin{center}
{\LARGE{\bf Avadh Saxena}} \\
{Theoretical Division and Center for Nonlinear Studies, Los
Alamos National Laboratory, Los Alamos, NM 87545, USA}
\end{center}

\vspace{.9in}
{\bf {Abstract}}

We obtain exact moving and stationary, spatially periodic and
localized solutions of a generalized discrete nonlinear
Schr\"odinger equation. More specifically, we find two different
moving periodic wave solutions and a localized moving pulse
solution. We also address the problem of finding exact stationary
solutions and, for a particular case of the model when stationary
solutions can be expressed through the Jacobi elliptic functions,
we present a two-point map from which all possible stationary
solutions can be found. Numerically we demonstrate the generic
stability of the stationary pulse solutions and also the
robustness of moving pulses in long-term dynamics.

\newpage

\section{Introduction}

The discrete nonlinear Schr\"odinger (DNLS) equation occurs
ubiquitously \cite{krb} throughout modern science. Most notable is
the role it plays in understanding the propagation of
electromagnetic waves in glass fibres and other optical waveguides
\cite{esm} as well as in the temporal evolution of Bose-Einstein
condensates \cite{tm}. One of the variants of the DNLS model is
the celebrated Ablowitz-Ladik (AL) model \cite{al} which is an
integrable model. Another aspect which stands out in favor of the
AL model is that, while most other discrete DNLS models have
stationary wave solutions \cite{krss}, this model has moving wave
solutions. Further, these moving waves avoid the discreteness
energy barrier (the so called Peierls-Nabarro (PN) barrier). These
solutions have played a major role in the computational studies of
the corresponding continuum NLS model \cite{ahs} as well as in
developing perturbation techniques \cite{kk}. It is clearly of
great interest to consider different variants of the DNLS equation
\cite{DNLSEx,pel,DNLSE1} and to try to obtain exact moving wave
solutions \cite{krss,DKYF}. The existence of such solutions
might help in discovering new integrable models and would also
help in further developing perturbative techniques in DNLS-type
equations. The purpose of this paper is to report on the existence
of exact moving as well as stationary solutions in a generalized
DNLS model with seven parameters. For finite lattices we find two
different periodic moving wave solutions while for the infinite
lattice we find a localized moving pulse solution.

In a recent paper, Pelinovsky \cite{pel}  has addressed the
question of spatial discretization of the NLS equation with cubic
nonlinearity
\be\label{1}
  i\dot{u} +u_{xx}+2|u|^2u =0\,.
\ee
While the {\it standard} choice for the DNLS equation is
\be\label{2}
  i\dot{u}_n +u_{n+1}+u_{n-1}-2u_{n}+2|u_n|^2u_n =0\,,
\ee
strictly speaking, there is no unique choice. Perhaps the only constraint
on the corresponding discrete model is that in
the continuum limit it should go over to the NLS Eq. (\ref{1}). By demanding
that the semi-discretization is symplectic and few other requirements,
Pelinovsky \cite{pel} showed that if one writes the DNLS equation in the form
\be\label{3}
  i\dot{u}_n +u_{n+1}+u_{n-1}-2u_{n}+f(u_{n-1},u_n,u_{n+1}) =0\,,
\ee
then the most general form for the nonlinear function $f$ is given by
\bea\label{4}
 &&f=\alpha_1|u_n|^2 u_n +\alpha_2|u_n|^2(u_{n+1}+u_{n-1})
 +\alpha_3u_n^2(\bar{u}_{n+1}+\bar{u}_{n-1}) \nonumber \\
 &&+\alpha_4 u_n(|u_{n+1}|^2+|u_{n-1}|^2)
 +\alpha_5 u_n(\bar{u}_{n+1}u_{n-1}+\bar{u}_{n-1}u_{n+1})
 \nonumber \\
 &&+\alpha_6 \bar{u}_n(u_{n+1}^2+u_{n-1}^2)
 +\alpha_7 \bar{u}_n u_{n+1}u_{n-1}
 +\alpha_8 (|u_{n+1}|^2 u_{n+1}+|u_{n-1}|^2 u_{n-1}) \nonumber \\
 &&+\alpha_9 (\bar{u}_{n-1}u_{n+1}^2+\bar{u}_{n+1}u_{n-1}^2)
 +\alpha_{10} (|u_{n+1}|^2 u_{n-1}+|u_{n-1}|^2 u_{n+1})\,,
 \eea
where $\bar{u}$ represents complex conjugate and the real valued
parameters ($\alpha_1,...,\alpha_{10}$) satisfy the continuity
constraint
 \be\label{5}
 \alpha_1+\alpha_7+2(\alpha_2+\alpha_3+\alpha_4+\alpha_5+\alpha_6+
 \alpha_8+\alpha_9+\alpha_{10})=2\,.
 \ee
The purpose of this paper is to obtain moving as well as stationary
solutions in this generalized model and study their stability.

We note in passing that, under weaker constraints than that used
in \cite{pel}, one can add to (\ref{4}) the term proportional to
$u_n(|u_{n-1} u_n|+|u_n u_{n+1}|)$, which was demonstrated to be
translationally invariant and conserving the norm, $\Sigma|u_n|^2$
\cite{DNLSEx}.

The paper is organized as follows. In Sec. \ref{Sec:Analytics} we
derive exact moving solutions for a seven-parameter DNLS model of
Eq. (\ref{4}) with $\alpha_1=\alpha_8=0$ under the constraint
(\ref{5}). In addition, for a five-parameter translationally
invariant DNLS equation we obtain a nonlinear map from which all
possible stationary solutions can be derived. In Sec.
\ref{Sec:Numerics} we present numerical results for the stationary
and moving pulse solutions to demonstrate their stability. Section
\ref{Sec:Concl} summarizes our main findings and concludes the
paper. In the Appendix we list the identities for the Jacobi
elliptic functions used in the derivation of the periodic wave
solutions.

\section{Analytical results} \label{Sec:Analytics}

We now show that two moving periodic wave solutions can be
obtained with this general cubic polynomial in case terms of the
type $|u_n|^2u_n$ are absent, i.e.
\be\label{6}
  \alpha_1=\alpha_8=0\,.
\ee
It may be added here that the famous AL moving wave solutions
are obtained in case only $\alpha_2$ is non-zero while all other
$\alpha_{i}$ are zero.

\subsection{ Solution I} \label{Sec:dn}

 In particular, it is not difficult to show that one of the
  exact periodic wave solution
 to Eq. (\ref{3}) [with $f$ being given by Eq. (\ref{4}) satisfying
 constraints (\ref{5}) and (\ref{6})] is given by
 \be\label{7}
 u_n = A \exp[-i(\omega t-kn+\delta)]\,\dn [\beta (n-v t+c),m]\,,
 \ee
 provided the following six relations are satisfied
 \be\label{8}
 v\beta=2A^2(\alpha_2-\alpha_3)\sin(k)\cs(\beta,m)\,,
 \ee
 \be\label{9}
 \cs(\beta,m)\alpha_6 \sin(2k)+[\alpha_9\sin(3k)-\alpha_{10}\sin(k)]
\cs(2\beta,m)=0\,,
 \ee
 \bea\label{10}
 &&\frac{\sin(k)}{A^2}=(\alpha_2-\alpha_3)\sin(k)\cs^2(\beta,m)
-\alpha_6\sin(2k)\ds(\beta,m)\ns(\beta,m) \nonumber \\
 &&-[\alpha_9\sin(3k)-\alpha_{10}\sin(k)]
[\cs^2(2\beta,m)+\ds(2\beta,m)\ns(2\beta,m)]\,,
 \eea
 \be\label{11}
 [\alpha_4+\alpha_6 \cos(2k)]\cs(\beta,m)
 +[\alpha_9\cos(3k)+\alpha_{10}\cos(k)]\cs(2\beta,m)=0\,,
 \ee
 \bea\label{12}
 &&\frac{\cos(k)}{A^2}=[\alpha_2+\alpha_3]\cos(k)\cs^2(\beta,m)
 -[\alpha_4+\alpha_6 \cos(2k)]\ds(\beta,m)\ns(\beta,m) \nonumber \\
 &&+[2\alpha_5\cos(2k)+\alpha_7]\cs(\beta,m)\cs(2\beta,m) \nonumber \\
 &&-[\alpha_9\cos(3k)+\alpha_{10}\cos(k)][\ds(2\beta,m)\ns(2\beta,m)
 -\cs^2(2\beta,m)]\,,
 \eea
 \bea\label{13}
 &&\frac{\omega}{A^2}-\frac{2}{A^2}=-2[\alpha_2+\alpha_3]
 \cos(k)\ds(\beta,m)\ns(\beta,m) \nonumber \\
 &&+2[\alpha_4+\alpha_6 \cos(2k)]\cs^2(\beta,m)
 -[2\alpha_5\cos(2k)+\alpha_7]\cs^2(\beta,m)\,.
 \eea
Here $c$ and $\delta$ are arbitrary constants, $k$, $\omega$, and
$v$ denote the wavenumber, frequency and velocity, respectively,
of the periodic wave whereas $\cs(a,m),\ds(a,m),\ns(a,m)$ stand for
the Jacobi elliptic functions $\cn(a,m)/\sn(a,m),
\dn(a,m)/\sn(a,m), 1/\sn(a,m)$ respectively with $m$ being the
modulus parameter ($0 \le m \le 1$) \cite{as}. While deriving
these relations, use has been made of the local identities
(\ref{14}) to (\ref{20}) for Jacobi elliptic functions $\dn(x,m)$
\cite{kls} which are given in the Appendix.

It may be noted that Eqs. (\ref{8}) to (\ref{13}) determine the
five parameters $A,\omega,k,v,\beta$ and give us one constraint
between the eight parameters $\alpha_2,...,\alpha_{10}$ (except
$\alpha_8$). In view of the constraint (\ref{5}) between these
parameters, it then follows that we have obtained a moving
periodic wave solution with six parameters. As expected, in the
limit $\alpha_2 \ne 0$ while all other $\alpha_i=0$, we recover
the well known periodic wave solution of the AL problem \cite{sb}.
Notice that in order that the periodic solution be compatible with
the lattice, the modulus $m$ has to be chosen such that $\beta
N_p=2K(m)$ where $K(m)$ denotes the complete elliptic integral of
the first kind \cite{as} and $N_p$ is the periodicity of the
lattice \cite{krss}.

\subsection{ Solution II} \label{Sec:cn}

As in the AL case, there is another periodic wave solution to the
DNLS Eq. (\ref{3}) with $f$ being given by Eq. (\ref{4})
satisfying constraints (\ref{5}) and (\ref{6}). It is given by
 \be\label{21}
 u_n = A \sqrt{m}\exp[-i(\omega t-kn+\delta)]\,\cn [\beta (n-v t+c),m]\,,
 \ee
 provided the following relations are satisfied
 \be\label{22}
 v\beta=2A^2(\alpha_2-\alpha_3)\sin(k)\ds(\beta,m)\,,
 \ee
 \be\label{23}
 \alpha_6 \sin(2k)\ds(\beta,m)+[\alpha_9\sin(3k)-\alpha_{10}\sin(k)]
\ds(2\beta,m)=0\,,
 \ee
 \bea\label{24}
 &&\frac{\sin(k)}{A^2}=(\alpha_2-\alpha_3)\sin(k)\ds^2(\beta,m)
-\alpha_6\sin(2k)\cs(\beta,m)\ns(\beta,m) \nonumber \\
&&-[\alpha_9\sin(3k)-\alpha_{10}\sin(k)]
[\ds^2(2\beta,m)+\cs(2\beta,m)\ns(2\beta,m)]\,,
 \eea
 \be\label{25}
 [\alpha_4+\alpha_6 \cos(2k)]\ds(\beta,m)
 +[\alpha_9\cos(3k)+\alpha_{10}\cos(k)]\ds(2\beta,m)=0\,,
 \ee
 \bea\label{26}
 &&\frac{\cos(k)}{A^2}=[\alpha_2+\alpha_3]\cos(k)\ds^2(\beta,m)
 -[\alpha_4+\alpha_6 \cos(2k)]\cs(\beta,m)\ns(\beta,m) \nonumber \\
 &&+[2\alpha_5\cos(2k)+\alpha_7]\ds(\beta,m)\ds(2\beta,m) \nonumber \\
 &&-[\alpha_9\cos(3k)+\alpha_{10}\cos(k)][\cs(2\beta,m)\ns(2\beta,m)
 -\ds^2(2\beta,m)]\,,
 \eea
 \bea\label{27}
 &&\frac{\omega}{A^2}-\frac{2}{A^2}=-2[\alpha_2+\alpha_3]
 \cos(k)\cs(\beta,m)\ns(\beta,m) \nonumber \\
&&+2[\alpha_4+\alpha_6 \cos(2k)]\ds^2(\beta,m)
 -[2\alpha_5\cos(2k)+\alpha_7]\ds^2(\beta,m)\,.
 \eea

\noindent While deriving these relations,
use has been made of the local identities (\ref{28}) to (\ref{34})
for the Jacobi elliptic function $\cn(x,m)$ \cite{kls} which have been
given in the Appendix.

As with the first solution, we again have a moving periodic wave
solution with six parameters and again in the limit when only
$\alpha_2 \ne 0$ while all other $\alpha_i$ are zero, we recover
the well known periodic wave solution of the AL problem \cite{sb}.
In addition, note that in order that the periodic solution be compatible
with the lattice, the modulus $m$ has to be chosen such that
$\beta N_p=4K(m)$ where $N_p$ is the periodicity of the lattice
\cite{krss}.

\subsection{ Two-Point maps for stationary solutions}
\label{Sec:TwoPointMap}

With the ansatz $u_n \left( t \right) = f_n e^{-i\omega t} $ we
obtain from the DNLS Eqs. (\ref{3}), (\ref{4}) the following
second-order difference equation for the amplitudes
\begin{eqnarray}\label{GenAmpl}
  f_{n - 1}  - (2 - \omega)f_n  + f_{n + 1} + \alpha_1 f_n^3
  + (\alpha _2  + \alpha _3)f_n^2 (f_{n - 1}  + f_{n + 1})
  + (\alpha _4  + \alpha _6)f_n
  \left( {f_{n - 1}^2  + f_{n + 1}^2 } \right) \nonumber \\
  +(2\alpha _5  + \alpha _7)f_{n - 1} f_n f_{n + 1}
  + \alpha_8 \left(f_{n - 1}^3 + f_{n + 1}^3 \right)
  + (\alpha_9  + \alpha_{10})f_{n - 1} f_{n + 1}(f_{n - 1} + f_{n + 1}) = 0.
\end{eqnarray}

For the following choice of parameters [that already includes the
continuity constraint (\ref{5})]
\be\label{stationary}
   \alpha_1=\alpha_8=0, \quad
   \alpha_4=-\alpha_6, \quad
   \alpha_9=-\alpha_{10},\, \quad {\rm and} \quad
   \alpha_7+2[\alpha_2+\alpha_3+\alpha_5]=2\,,
\ee
we get from (\ref{GenAmpl}) the following second-order
difference equation for the amplitudes
\begin{eqnarray}\label{amplitudes}
  f_{n - 1}  - \left( {2 - \omega } \right)f_n  + f_{n + 1}
  + \left( {\alpha _2  + \alpha _3 } \right)f_n^2
  \left( {f_{n - 1}  + f_{n + 1} } \right)
  + \left( {2\alpha _5  + \alpha _7 } \right)f_{n - 1} f_n f_{n + 1} = 0.
\end{eqnarray}
In this case, the
stationary problem is exactly solvable. Indeed, one can obtain
the first integral of (\ref{amplitudes}) and present it in the
form of a two-point nonlinear map
\begin{eqnarray}\label{map}
   f_{n + 1} &=& (2 - \omega)\frac{Zf_n \pm \sqrt {R(f_n)} }
   {{2 - \omega  + Yf_n^2 }}, \nonumber \\
   R(f_n)&=&-\frac{Y}
   {2 - \omega} (K - X f_n^2 + f_n^4) ,
\end{eqnarray}
where
\begin{eqnarray}\label{mapparam}
  Z = \frac{{(2 - \omega)^2
  - K\left( {2\alpha _5  + \alpha _7 } \right)^2 }}
  {{2K\left( {\alpha _2  + \alpha _3 } \right)\left( {2\alpha _5  + \alpha _7 } \right)
  + 2\left( {2 - \omega } \right)}}, \nonumber \\
  Y = 2\left( {\alpha _2  + \alpha _3 } \right)Z
  + \left( {2\alpha _5  + \alpha _7 } \right), \nonumber \\
  X = -\frac{KY^2 + (2 - \omega)^2(1 - Z^2)}{(2 - \omega)Y}.
\end{eqnarray}
Apart from the model parameters $\alpha_i$ and frequency $\omega$,
the nonlinear map (\ref{map}), (\ref{mapparam}) contains the
integration constant $K$. Due to the symmetry of equation
(\ref{amplitudes}) one can substitute $f_{n+1}$ for $f_{n-1}$ in
(\ref{map}). For any set of admissible values $f_0$, $K$, and
$\omega$ one can find the amplitudes of a stationary solution
by iterating (\ref{map}). For $R(f_n)>0$ the map (\ref{map}) gives
two values for $f_{n+1}$ and one should take the one which
satisfies the original three-point problem (\ref{amplitudes}). It
is sufficient to take $f_{n+1}$ different from $f_{n-1}$.

The above two-point map can also be constructed from the Jacobi
elliptic function solutions (\ref{7}) or (\ref{21}) as described
in our recent work on a discrete $\phi^4$ model
\cite{DKKSinpress}. The corresponding DNLS equation has five free
parameters because (\ref{stationary}) sets up five constraints
between the ten parameters ($\alpha_i$) of the model. We note that
{\em any} stationary solution to the DNLS equation defined by
(\ref{3}) and (\ref{4}) with the parameters satisfying
(\ref{stationary}) can be constructed from the nonlinear map
(\ref{map}), (\ref{mapparam}). Such investigations have been
carried out in our recent work on the DNLS equation \cite{DKYF}
and the $\phi^4$ equation \cite{DKKSinpress,DKYF2006PRE}.

It is also worth pointing out that the three-point problem given by
Eq. (\ref{amplitudes}) and the three-point problem studied by Quispel
{\it et al.} \cite{Quispel} both can be presented in the following
general form
\be\label{qui}
   f_{n+1}=\frac{h_1(f_n)-h_2(f_n)f_{n-1}}{h_2(f_n)-h_3(f_n)f_{n-1}}\,.
\ee
For a particular choice of the functions $h_i(f_n)$, Quispel {\it
et al.} have found a two-point map (i.e., the first integral of the
corresponding three-point problem) which is quadratic in both
$\phi_n$ and $\phi_{n+1}$ \cite{Quispel}. For our choice of these
functions,
\be\label{qui1}
   h_1(f_n)=(2-\omega)f_n\,,~~~h_2(f_n)=1+(\alpha_2+\alpha_3)f_n^2\,,
   ~~~h_3(f_n)=-(2\alpha_5+\alpha_7)f_n\,,
\ee
we found the map (\ref{map}) which is, in general, quartic
in $\phi_n$ and quadratic in $\phi_{n+1}$. Clearly, our map
(\ref{map}) does not belong to the 12 parameter map discussed in
\cite{Quispel}.

The above result is new in that it generalizes the map reported
in our recent work \cite{DKKSinpress}. For completeness, let us
also reproduce here the well-known result \cite{pel,DKKSinpress}
for the case of
\be\label{stationary2}
   \alpha _8  = \alpha _9  + \alpha _{10}, \quad
   \alpha_1 =\alpha_4  + \alpha_6, \quad
   \alpha _1  = 2\alpha_5  + \alpha_7\,, \quad {\rm and}
   \quad 4\alpha_1 +2[\alpha_2+\alpha_3 + 2\alpha_8]=2\,,
\ee
when the continuity constraint (\ref{5}) is satisfied and
(\ref{GenAmpl}) reduces to the following second-order
difference equation
\begin{eqnarray}\label{Add1}
  f_{n - 1}  - (2 - \omega)f_n  + f_{n + 1}
  + \alpha_1 f_n \left[
   f_{n - 1}^2 + f_n^2 + f_{n + 1}^2
  +f_{n - 1}f_{n + 1} \right] \nonumber \\
  + (\alpha _2  + \alpha _3)f_n^2 (f_{n - 1}  + f_{n + 1})
  + \alpha_8 \left[f_{n - 1}^3 + f_{n + 1}^3
  + f_{n - 1} f_{n + 1}(f_{n - 1} + f_{n + 1}) \right] = 0.
\end{eqnarray}
The first integral of (\ref{Add1}) is
\begin{eqnarray}\label{Add2}
  V\left( {f_{n - 1},f_n} \right) \equiv f_{n - 1}^2
  + f_n^2  - \left( {2 - \omega }
  \right)f_{n - 1}f_n + \alpha _1 \left( {f_{n - 1}^2  + f_n^2 } \right)
  f_{n - 1}f_n \nonumber \\ + \left(
  {\alpha _2  + \alpha _3 } \right)f_{n - 1}^2 f_n^2
  + \alpha _8 \left( {f_{n - 1}^4  + f_n^4 } \right) + K = 0,
\end{eqnarray}
where $K$ is the integration constant. This is so because
(\ref{Add1}) can be rewritten in the form
\begin{eqnarray}\label{Add3}
  \frac{V(f_n,f_{n + 1})- V(f_{n-1},f_n)}{f_{n + 1} -f_{n - 1}} = 0,
\end{eqnarray}
and one can verify that if $V\left( {f_{n - 1},f_n} \right)=0$ then
(\ref{Add1}) is satisfied. Solving the algebraic problem
(\ref{Add2}) iteratively for an admissible initial value $f_0$ one
can construct a stationary solution to (\ref{Add1}). This
model has six free parameters because (\ref{stationary2}) sets up
four constraints between the ten parameters ($\alpha_i$) of the model.
In general, stationary solutions to the DNLS equation with the
parameters satisfying (\ref{stationary2}) cannot be expressed in
terms of the Jacobi elliptic functions, but, as it was already
mentioned, they can be constructed iteratively from (\ref{Add2})
and they can be placed anywhere with respect to the lattice sites.

We note that the translationally invariant discrete models
possessing the form of equation (\ref{Add3}) have been introduced
by Kevrekidis in \cite{Kevrekidis:2003-68:PD}.

\subsection{Moving and stationary pulse solution} \label{Sec:Pulse}

In the limit $m \rightarrow 1$, both the periodic moving wave
solutions (\ref{7}) and (\ref{21}) reduce to the localized moving
pulse solution
 \be\label{35}
u_n = A \exp[-i(\omega t-kn+\delta)]\,\sech [\beta (n-v t+c)]\,,
 \ee
 and the relations (\ref{8}) to (\ref{13}) [as well as (\ref{22}) to
(\ref{27})] take a simpler form
 \be\label{36}
 v=\frac{2\sin(k)\sinh(\beta)}{\beta}\,,
 \ee
 \be\label{37}
2\alpha _6 \sin(2k)\cosh (\beta)  + \alpha _9 \sin(3k) - \alpha
_{10} \sin (k) = 0\,,
 \ee
 \be\label{38}
 \left[ {\sinh ^2 (\beta)  + \left( {\alpha _3  - \alpha _2 }
\right)A^2 } \right]\sin (k) = 0\,, \ee
 \be\label{39}
 A^2  = \frac{{2\sinh ^2 (\beta) \cosh (\beta) \cos (k)}}{{2\left(
{\alpha _2  + \alpha _3 } \right)\cos (k)\cosh (\beta)  + 2\left(
{\alpha _5  - \alpha _6 } \right)\cos(2 k) + \alpha _7 - 2\alpha
_4 }}\,,
 \ee
 \be\label{40}
2[\alpha _4 + \alpha _6\cos(2k)] \cosh (\beta) + \alpha _9 \cos
(3k) + \alpha _{10} \cos (k) = 0\,,
 \ee
 \be\label{41}
 \omega=2[1-\cos(k)\cosh(\beta)]\,.
 \ee

From (\ref{36}), pulse velocity is zero when $k=0$ or $k=\pi$. In
the former case we have the non-staggered stationary pulse while
in the latter case we have the staggered pulse. Needless to say
that these remarks equally apply to the periodic wave solutions
(\ref{7}) and (\ref{21}). In particular for $k=0$ we obtain the
non-staggered stationary pulse solution
\begin{eqnarray}\label{NonStaggStationaryPulse}
   u_n = A \exp[-i(\omega t+\delta)]\,\sech [\beta (n+c)]\,,
   \nonumber \\
   \omega  = 2 - 2\cosh (\beta) , \nonumber \\
   A^2  = \frac{{2\sinh ^2 (\beta) \cosh (\beta) }}
   {{2\left( {\alpha _2  + \alpha _3 } \right)\cosh (\beta)
   + 2\left( {\alpha _5  - \alpha _4  - \alpha _6 } \right) + \alpha _7 }}, \nonumber \\
   2\left( {\alpha _4  + \alpha _6 } \right)\cosh (\beta)  + \alpha _9  + \alpha _{10}  = 0,
\end{eqnarray}
while in the latter case we obtain the staggered stationary pulse,
\begin{eqnarray}\label{StaggStationaryPulse}
   u_n = (-1)^n A \exp[-i(\omega t+\delta)]\,\sech [\beta (n+c)]\,,
   \nonumber \\
   \omega  = 2 + 2\cosh (\beta) , \nonumber \\
   A^2  = \frac{{ - 2\sinh ^2 (\beta) \cosh (\beta) }}
   {{ - 2\left( {\alpha _2  + \alpha _3 } \right)\cosh (\beta)
   + 2\left( {\alpha _5 - \alpha _4 - \alpha _6 } \right) + \alpha _7
   }}, \nonumber \\
   2\left( {\alpha _4  + \alpha _6 } \right)\cosh (\beta)
   - \alpha _9  - \alpha _{10}  = 0.
\end{eqnarray}

\begin{figure}
\includegraphics{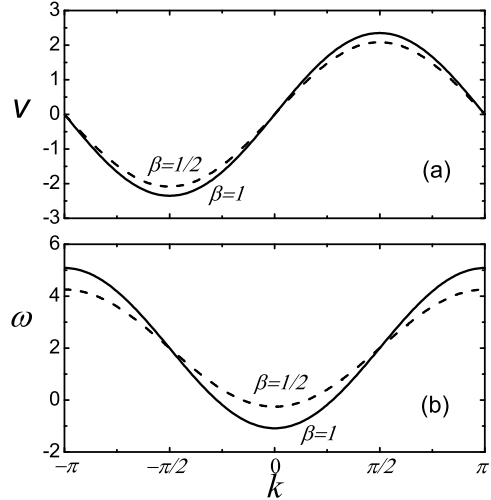}
\caption{(a) Velocity $v$ and (b) frequency $\omega$ of the pulse
as functions of the wavenumber parameter $k$ at fixed value of the
other parameter, inverse width of the pulse, $\beta=1/2$ (dashed
lines) and $\beta=1$ (solid lines).  These functions are defined
by (\ref{36}) and (\ref{41}) and they do not depend on the model
parameters $\alpha_i$. Pulse velocity is zero at $k=0$ and
$k=\pi$, the former case corresponds to the non-staggered
stationary pulse (\ref{NonStaggStationaryPulse}) while the latter
case to the staggered stationary pulse
(\ref{StaggStationaryPulse}).} \label{Figure1}
\end{figure}

\begin{figure}
\includegraphics{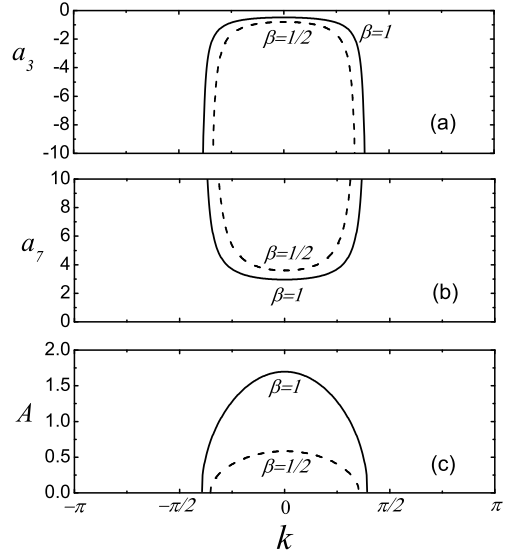}
\caption{Nonzero model parameters $\alpha_3$ and $\alpha_7$ and
the pulse amplitude $A$ as functions of the parameter $k$ at fixed
value of the other parameter $\beta=1/2$ (dashed lines) and
$\beta=1$ (solid lines). These functions are defined by
(\ref{38a}). For $\beta=1/2$ the solution exists (i.e. $A$ is
real) for $|k|< 1.11$ while for $\beta=1$ it exists for $|k|<
1.23$. The velocity $v$ and frequency $\omega$ of the pulse are
shown in Fig. \ref{Figure1}.} \label{Figure2}
\end{figure}

\begin{figure}
\includegraphics{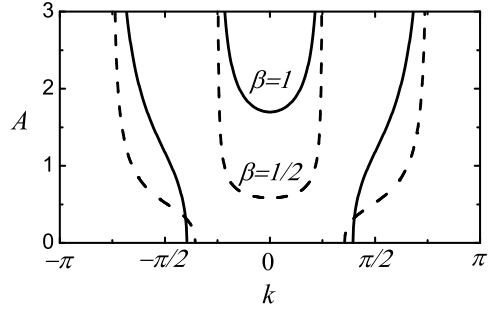}
\caption{Pulse amplitude $A$ as a function of the parameter $k$ at
$\beta=1/2$ (dashed lines) and $\beta=1$ (solid lines) for the
model with three nonzero parameters, $\alpha_3$, $\alpha_5$ and
$\alpha_7$. Here we set $\alpha_5=1$ and find other model and
pulse parameters from (\ref{38b}). For $\beta=1/2$ the solution
exists (i.e. $A$ is real) for $|k|< \pi/4$ and $1.12<|k|< 3\pi/4$,
while for $\beta=1$ it exists for $|k|< \pi/4$ and $1.24<|k|<
3\pi/4$. The velocity $v$ and frequency $\omega$ of the pulse are
shown in Fig. \ref{Figure1}.} \label{Figure3}
\end{figure}

\begin{figure}
\includegraphics{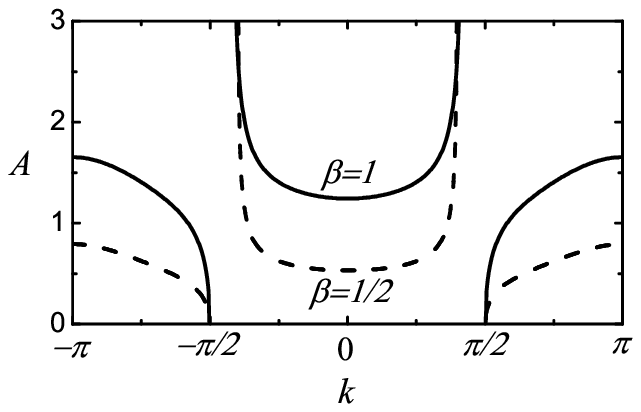}
\caption{Pulse amplitude $A$ as a function of the parameter $k$ at
$\beta=1/2$ (dashed lines) and $\beta=1$ (solid lines) for the
model with three nonzero parameters, $\alpha_2$, $\alpha_3$ and
$\alpha_5$. The relation between model and pulse parameters are given
by (\ref{38c}). For $\beta=1/2$ the solution exists (i.e. $A$ is
real) for $|k|< 1.26$ and $\pi/2 <|k|\le \pi$, while for $\beta=1$
it exists for $|k|< 1.32$ and $\pi/2 <|k|\le \pi$. The velocity $v$
and frequency $\omega$ of the pulse are shown in Fig.
\ref{Figure1}.} \label{Figure4}
\end{figure}

\section{ Analysis of the pulse solution and numerical results}
\label{Sec:Numerics}

For given model parameters $\alpha_i$, the moving pulse solution
(\ref{35})-(\ref{41})
is characterized by two parameters, $\beta>0$ and $-\pi < k \le
\pi$. As it can be seen from (\ref{36}) and (\ref{41}), the pulse
velocity and frequency do not depend on model parameters while the
pulse amplitude does, see (\ref{39}). Using (\ref{36}) one can
express $\omega$ in (\ref{41}) as function of $v$ and $\beta$.
Also using (\ref{41}) one can express the group velocity
$d\omega/dk$. The pulse solution exists for given $\beta$ and $k$
if the right-hand side of (\ref{39}) is positive and if
(\ref{37}), (\ref{38}), and (\ref{40}) can be satisfied together
with the continuity constraint (\ref{5}), where we assume
(\ref{6}).

As for the stationary pulse solution
(\ref{NonStaggStationaryPulse}) or (\ref{StaggStationaryPulse}),
for given model parameters $\alpha_i$, the moving pulse solution
is characterized by a single parameter $\beta>0$. In general, as
far as the model parameters are fixed, the parameter $\beta$ of
the stationary pulse is also fixed through the last equation in
(\ref{NonStaggStationaryPulse}) or (\ref{StaggStationaryPulse}).
However, for $\alpha _4 = -\alpha _6$ and $\alpha _9 = -\alpha
_{10}$, this constraint disappears and $\beta$ can change
continuously within a domain where $A^2>0$.
Recall that in this particular case
the stationary pulse solution can also be constructed from the
two-point map presented in Sec. \ref{Sec:TwoPointMap}, for which
one should set the integration constant $K=0$.

\subsection{Different moving solutions}

Coming back to the moving pulse solution (\ref{35})-(\ref{41}),
several comments are in order.

\begin{enumerate}

\item The relations (\ref{36}) and (\ref{41}) are exactly the same as in the
AL case \cite{al}. It is indeed remarkable that the velocity $v$
and the frequency $\omega$ in our case are identical to those in
the AL model even though our model has eight nonlinear terms (with
coefficients $\alpha_2$ to $\alpha_{10}$ with $\alpha_8=0$) while
AL has only one term with $\alpha_2=1$.
It is amusing to
note that these two relations have also been obtained by
Pelinovsky and Rothos from an entirely different approach
\cite{pr}, namely from the linear dispersion relation for the
corresponding differential advance-delay equation. In Fig.
\ref{Figure1} we show how $v$ and $\omega$ depend on one of the
pulse parameter, $k$, at fixed values of the other parameter,
$\beta=1/2$ (dashed lines) and $\beta=1$ (solid lines).

\item Unfortunately, we do not know the Hamiltonian from which the
DNLS Eq. (\ref{3}) with $f$ given by Eq. (\ref{4}) can be derived.
As a result, we cannot demonstrate the absence of the PN barrier
from the energy consideration. However, since our stationary
solutions have an effective translational invariance (i.e. the
solution is valid for any value of the constant $c$), this
suggests that the PN barrier would be zero for these solutions.

\item From Eq. (\ref{38}) it follows that the moving pulse solution
exists only if $\alpha_2$ and/or $\alpha_3$ are nonzero.
Further, in case $\alpha_2=0$, then it follows from Eq. (\ref{38}) that
$\alpha_3 < 0$.

\item In the limit when only $\alpha_2$ is nonzero while all other
$\alpha_i$ are zero, we recover the well known AL moving pulse
solution \cite{al}.

\item The $\sn$-type and hence dark soliton solution can also be obtained
in this generalized model provided the right hand side of the continuity
Eq. (\ref{5}) is $-2$ (instead of 2).

\item In case only $\alpha_2$ and/or $\alpha_3$ are nonzero while all other
$\alpha_i=0$, then the generalized DNLS Eq. (\ref{3}) with $f$
given by Eq. (\ref{4}) conserves the momentum defined by
\be\label{mom1} P=i\sum_{n} \big
(u_{n+1}\bar{u}_n-\bar{u}_{n+1}u_n \big ) \,. \ee On the other
hand, in case only $\alpha_5$ and/or $\alpha_7$ are nonzero while
all other $\alpha_i=0$, then the generalized DNLS Eq. (\ref{3})
with $f$ given by Eq. (\ref{4}) conserves the momentum defined by
\be\label{mom2} P=i\sum_{n} \big
(u_{n+2}\bar{u}_n-\bar{u}_{n+2}u_n \big ) \,. \ee
Expression (\ref{mom2}) is similar to that introduced in
\cite{DKKSinpress} for the $\phi^4$ discrete equation.

\item From Eqs. (\ref{36}) to (\ref{41}) it follows that the moving
pulse solution is also possible when only two of the eight
parameters are nonzero. For example, the moving pulse solution
(\ref{35}) exists in case $\alpha_3,\alpha_7$ are nonzero while
all other $\alpha_i$ are zero. While the relations (\ref{36}) and
(\ref{41}) are always valid, the other relations and the
constraint (\ref{5}) take the form
\be\label{38a}
   \alpha _3  = \frac{1}{{1 - 2\cos (k)\cosh (\beta)}}, \quad
   \alpha _7  = 2\left( {1 - \alpha _3 }\right), \quad
   A^2  = \frac{{\cos (k)\cosh (\beta) \sinh ^2 (\beta)}}
   {{\alpha _3 [\cos (k)\cosh (\beta) - 1]  + 1}}.
\ee
For a pair of pulse parameters, $k$ and $\beta$, we find
$\alpha_3$ and then $\alpha_7$ and $A$ from (\ref{38a}) and
present the result in Fig. \ref{Figure2} for $\beta=1/2$ (dashed
lines) and $\beta=1$ (solid lines). For $\beta=1/2$ the solution
exists (i.e. $A$ is real) for $|k|< 1.11$ while for $\beta=1$ it
exists for $|k|< 1.23$. One can see that the non-staggered
stationary pulse ($k=0$) exists while staggered stationary pulse
($k=\pi$) does not exist in this case.

\item The moving pulse solution (\ref{35}) also exists in case only (i)
$\alpha_3,\alpha_5$ are nonzero; (ii) $\alpha_2,\alpha_5$ are
nonzero and $\cos(2k)=0$, i.e., regardless of the model parameters,
in this case one can have only $k=\pm \pi/4$ and $k=\pm 3\pi/4$.
Constraints similar to those in (\ref{38a}) are easily written
down from relations (\ref{5}) and (\ref{36}) to (\ref{41}). We
were unable to find other sets of model parameters supporting the
pulse solution when there are only two nonzero parameters.

\item There are several possibilities, with three of the eight $\alpha_i$
being nonzero (the remaining five $\alpha_i$ being zero), in which case
the moving pulse solution (\ref{35}) is still valid. These
cases are: (i) $\alpha_2, \alpha_3, \alpha_5$ are nonzero; (ii)
$\alpha_2, \alpha_3, \alpha_7$ are nonzero; (iii) $\alpha_3,
\alpha_5, \alpha_7$ are nonzero; (iv) $\alpha_2, \alpha_5,
\alpha_7$ are nonzero; (v) $\alpha_2, \alpha_4, \alpha_6$ are
nonzero with $\alpha_4=\alpha_6$ and $k=\pm\pi/2$; (vi) $\alpha_3,
\alpha_4, \alpha_6$ are nonzero with $\alpha_4=\alpha_6$ and
$k=\pm\pi/2$; (vii) $\alpha_2, \alpha_9, \alpha_{10}$ are nonzero
with $\alpha_9=\alpha_{10}$ and $k=\pm\pi/4$ or $k=\pm 3\pi/4$.

In all these cases the constraints similar to those in (\ref{38a})
are easily obtained from relations (\ref{5}) and (\ref{36}) to
(\ref{41}). For example, in case only $\alpha_3, \alpha_5,
\alpha_7$ are nonzero, while the relations (\ref{36}) and
(\ref{41}) are always valid, the other relations and the
constraint (\ref{5}) take the form
\be\label{38b}
   \alpha _3  = \frac{{1 - 2\alpha _5 \sin ^2 \left( k \right)}}
   {{1 -2\cos \left( k \right)\cosh \left( \beta \right)}}, \quad
   \alpha_7 = 2\left( {1 - \alpha _3  - \alpha _5 } \right), \quad
   A^2  = \frac{{\cos \left( k \right)\sinh ^2 (\beta)\cosh
   \left( \beta  \right)}}{{1 + \alpha _3 \left[ {\cos \left( k
   \right)\cosh \left( \beta  \right) - 1} \right] - 2\alpha _5 \sin
   ^2 \left( k \right)}}.
\ee
The number of constraints in this case is such that one has a free
model parameter, say $\alpha_5$, and pulse parameters $k$ and
$\beta$ can change continuously within a certain domain. For
$\alpha_5$ with a small absolute value the solution is close to
(\ref{38a}) shown in Fig. \ref{Figure2}, but, for example, for
$\alpha_5=1$ the result is qualitatively different, as it can be
seen from Fig. \ref{Figure3}. Also note that in this case the
non-staggered stationary pulse ($k=0$) exists while the staggered
stationary pulse ($k=\pi$) does not exist.

On the other hand, in case only $\alpha_2, \alpha_3, \alpha_5$ are
nonzero we have the following constraints
\begin{eqnarray}\label{38c}
   \alpha_3 = \frac{-\alpha_5\cos(2k)}{2\cos(k)\cosh(\beta)}, \quad
   \alpha _2 = 1 - \alpha _3- \alpha _5 ,
   \quad A^2 = \frac{{\cos(k)\sinh ^2(\beta)\cosh(\beta)}}
   {{\left( {\alpha _2  + \alpha _3 } \right)\cos(k)\cosh
   (\beta) + \alpha _5 \cos (2 k)}}.
\end{eqnarray}
The relation between pulse parameters and model parameters in this
case are shown in Fig. \ref{Figure4}. In this case one has both
non-staggered and staggered stationary pulse solutions for $k=0$
and $k=\pi$, respectively.

We give two more solutions, for the case when only $\alpha_2,
\alpha_4, \alpha_6$ are nonzero,
\begin{eqnarray}\label{38d}
   A^2  = \frac{{\sinh ^2 \beta }}{{\alpha _2 }},
   \quad \alpha _4  =\frac{1 - \alpha _2 }{2},
   \quad \alpha _4  = \alpha _6, \quad k = \pm \frac{\pi}{2},
\end{eqnarray}
and for the case when only $\alpha_2, \alpha_9, \alpha_{10}$ are
nonzero,
\begin{eqnarray}\label{38e}
   A^2  = \frac{{\sinh ^2 \beta }}{{\alpha _2 }},
   \quad \alpha _{10}= \frac{{1 - \alpha _2 }}{2},
   \quad \alpha _9  = \alpha _{10},
   \quad k = \pm \frac{{ \pi }}{4}, \quad {\rm or}
   \quad k = \pm \frac{{ 3\pi }}{4}.
\end{eqnarray}
These two {\em moving} solutions are interesting because for them
the relations (\ref{stationary}) are violated.
These models have one free parameter, for example, $\alpha_2 >0$.
Among the two pulse parameters, only $\beta$ can change
continuously, while $k$ can assume only a few isolated discrete
values, that do not depend on model parameters $\alpha_i$. For the
cases when there are only three nonzero parameters, we were unable
to find sets of model parameters supporting the pulse solution
other than the ones described above.

\item Similarly, there are several possibilities when less than eight
parameters are nonzero and still the moving pulse solution
(\ref{35}) continues to exist and relations similar to those in
Eq. (\ref{38a}) can easily be obtained in all these cases.

\item Since the DNLS equation (\ref{3}), (\ref{4}) with any set
of parameters $\alpha_i$ satisfying the continuity constraint
(\ref{5}) reduces to the same NLS equation (\ref{1}), for a
sufficiently wide (small $\beta$) and slow (small $|k|$) pulse,
all the solutions given above are close and can be well
approximated by the moving solution to the continuous NLS
equation.

\end{enumerate}

\subsection{Stability of the pulse solution}

Let us now discuss the small amplitude vibration spectrum for the
lattice containing a stationary pulse in order to observe the
peculiarities of the spectrum of the pulse in a translationally
invariant lattice and to discuss the stability of the pulse. The
vibrational spectrum was calculated following the methodology
presented in Ref. \cite{Carr:1985-201:PLA} similar to the work
in \cite{DKYF}. In brief, we consider a small
complex perturbation of a stationary solution and substitute the
ansatz $u_n \left( t \right) = [f_n +\varepsilon_n(t)] e^{-i\omega
t} $ with $\varepsilon_n(t) =a_n(t)+ ib_n(t)$ into the DNLS
equation (\ref{3}), (\ref{4}) and obtain a linear equation for
$\varepsilon_n(t)$. Separating real and imaginary parts of the
equation, we derive the following system
\begin{eqnarray}
  \left( {\begin{array}{*{20}c}
    {{\mathbf{\dot b}}}  \\
    {{\mathbf{\dot a}}}  \\
  \end{array} } \right)
  = \left( {\begin{array}{*{20}c}
    0 & \mathbf{K}  \\
    \mathbf{J} & 0  \\
  \end{array} } \right)
  \left( {\begin{array}{*{20}c}
    {\mathbf{b}}  \\
    {\mathbf{a}}  \\
  \end{array} } \right),
  \label{EigProblem}
\end{eqnarray}
where vectors $\mathbf{a}$ and $\mathbf{b}$ contain $a_n$ and
$b_n$, respectively, while the nonzero coefficients of matrices
$\mathbf{K}$ and $\mathbf{J}$ are given by,
\begin{eqnarray}\label{K}
   K_{n,n-1}  = 1 + \left( {\alpha _2  + \alpha _3 } \right)f_n^2
   + \left( {2\alpha _5  + \alpha _7 } \right)f_n f_{n + 1}
   + (\alpha_9 + \alpha_{10})\left(f_{n+1}^2 + 2f_{n-1} f_{n+1}\right) , \nonumber \\
   K_{n,n}  =  - \left( {2 - \omega } \right) + 2\left( {\alpha _2
   + \alpha _3 } \right)f_n \left( {f_{n - 1}  + f_{n + 1} } \right)
   + (\alpha_4 + \alpha_6)\left(f_{n-1}^2 + f_{n+1}^2\right)
   + \left( {2\alpha _5  + \alpha _7 } \right)f_{n - 1} f_{n + 1}, \nonumber \\
   K_{n,n+1}  = 1 + \left( {\alpha _2  + \alpha _3 } \right)f_n^2
   + \left( {2\alpha _5  + \alpha _7 } \right)f_{n - 1} f_n
   + (\alpha_9 + \alpha_{10})\left(f_{n - 1}^2 + 2f_{n - 1} f_{n + 1}\right) ,
\end{eqnarray}
\begin{eqnarray}\label{J}
   J_{n,n - 1}  =  - 1 - (\alpha _2  - \alpha _3)f_n^2
   - 2\alpha _6 f_{n - 1} f_n  - \alpha _7 f_n f_{n + 1}
   - 2\alpha _9 f_{n - 1} f_{n + 1}  + \left( {\alpha _9
   - \alpha _{10} } \right)f_{n + 1}^2,  \nonumber \\
   J_{n,n}  = \left( {2 - \omega } \right) - 2\alpha _3 f_n
   \left( {f_{n - 1}  + f_{n + 1} } \right) - \left( {\alpha _4
   - \alpha _6 } \right)\left( {f_{n - 1}^2  + f_{n + 1}^2 } \right)
   - \left( {2\alpha _5  - \alpha _7 } \right)f_{n - 1} f_{n + 1}, \nonumber \\
   J_{n,n + 1}  =  - 1 - \left( {\alpha _2  - \alpha _3 } \right)f_n^2
   - 2\alpha _6 f_n f_{n + 1}  - \alpha _7 f_{n - 1} f_n  - 2\alpha _9
   f_{n - 1} f_{n + 1}  + (\alpha_9 - \alpha_{10})f_{n - 1}^2.
\end{eqnarray}
A stationary solution is characterized as linearly stable if and
only if the eigenvalue problem
\begin{eqnarray} \label{EigValProblem}
   \left( {\begin{array}{*{20}c}
   0 & {\mathbf{K }}  \\    {\mathbf{J}} & 0  \\
   \end{array} } \right) \left( {\begin{array}{*{20}c} {\mathbf{b}}  \\
   {\mathbf{a}} \\ \end{array} } \right)= \gamma
   \left( {\begin{array}{*{20}c} {\mathbf{b}}  \\
   {\mathbf{a}} \\ \end{array} } \right)
\end{eqnarray}
results in nonpositive real parts of all eigenvalues $\gamma$.

Setting in the above matrices $f_n=0$, and solving the resulting
eigenvalue problem one finds the spectrum of vacuum
\begin{eqnarray}\label{vacuum}
   \Omega  =  \pm \left[-\omega + 4\sin^2 \left(\frac{Q}{2}\right) \right],
\end{eqnarray}
where $\Omega$ and $Q$ are the frequency and the wavenumber of a
small-amplitude harmonic mode, respectively.
A stationary pulse was placed in the middle of a lattice of $N=200$
points and the eigenvalue problem (\ref{EigValProblem}) was solved
employing periodic boundary conditions.
Here we do not aim to present a comprehensive numerical study of
the stability of the pulse because the DNLS equation under
consideration has a multi-dimensional parameter space and such an
exhaustive study would entail enormous effort. Instead, our intent
is to check several sets of parameters and to provide a few examples
illustrating the generic
stability of the pulse solution.

\begin{figure}
\includegraphics{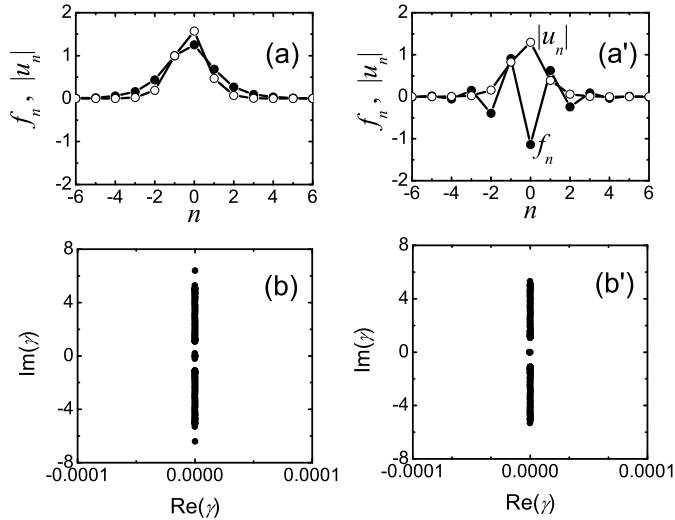}
\caption{Two examples of (a), (a') stationary pulse profiles and
(b), (b') their spectra. Left panels show the results for a
non-staggered pulse, while right panels are for a staggered pulse.
Model parameters correspond to a translationally invariant lattice,
i.e., they satisfy (\ref{stationary}). For (a), (b) parameters are
$\alpha_2=1$, $\alpha_3=-1/2$, $\alpha_4=-\alpha_6=1/2$,
$\alpha_5=-1/2$, $\alpha_7=2$, and $\alpha_9=-\alpha_{10}=-1/2$.
For (a'), (b') parameters are $\alpha_2=1$, $\alpha_3=0$,
$\alpha_4=-\alpha_6=1/2$, $\alpha_5=-1/2$, $\alpha_7=1$, and
$\alpha_9=-\alpha_{10}=-1/2$. The pulses are defined by,
respectively, (\ref{NonStaggStationaryPulse}) and
(\ref{StaggStationaryPulse}) with parameters $\beta=1$,
$\delta=0$, and $c=0.25$. Pulses are placed asymmetrically with
respect to the lattice, nevertheless, they are stationary and
stable since all eigenvalues $\gamma$ have zero real parts. The
spectra also contain two pairs of zero eigenvalues, one pair
corresponds to the translational invariance and another to the
invariance with respect to the phase shift.} \label{Figure5}
\end{figure}

Two examples of stationary, stable pulses and their spectra are
presented in Fig. \ref{Figure5}. Left panels present the results
for a non-staggered pule, while the right panels are for a staggered
pulse. Model parameters correspond to a translationally invariant
lattice, i.e., they satisfy (\ref{stationary}). For panels (a), (b)
parameters are $\alpha_2=1$, $\alpha_3=-1/2$, $\alpha_4=-\alpha_6=1/2$,
$\alpha_5=-1/2$, $\alpha_7=2$, and $\alpha_9=-\alpha_{10}=-1/2$.
For panels (a'), (b') parameters are $\alpha_2=1$, $\alpha_3=0$,
$\alpha_4=-\alpha_6=1/2$, $\alpha_5=-1/2$, $\alpha_7=1$, and
$\alpha_9=-\alpha_{10}=-1/2$. The choice of parameters is rather
arbitrary. For the non-staggered pulse all coefficients are
nonzero so that all terms of the DNLS equation are involved. For
the staggered pulse we found that nonzero $\alpha_3$ makes the
pulse unstable, that is why we set this coefficient equal to zero.
The non-staggered and staggered pulses are defined by,
respectively, (\ref{NonStaggStationaryPulse}) and
(\ref{StaggStationaryPulse}) with parameters $\beta=1$,
$\delta=0$, and $c=0.25$. We then found for the non-staggered pulse
$\omega=-1.0862$ and $A=1.2946$, and for the staggered one,
$\omega=5.0862$ and $A=1.1752$. The pulses are placed asymmetrically
with respect to the lattice, nevertheless, they are stationary and
stable since all eigenvalues $\gamma$ have zero real parts. The
spectrum of non-staggered pulse contains the spectrum of vacuum
(\ref{vacuum}) with the bands $1.0862 \le |\Omega| \le 5.0862$;
the three pulse internal modes with frequencies $\pm 0.195$, $\pm
5.290$, and $\pm 6.396$; and the two pairs of zero eigenvalues,
one pair corresponding to the translational invariance and another
to the invariance with respect to the phase shift. The spectrum of
staggered pulse is similar but it contains not three but only one
pulse internal mode with frequencies $\pm 5.308$.

We have checked the stability of stationary pulses (both non-staggered
and staggered) with different $\beta \sim 1$, and also different
positions with respect to the lattice, $c$, and for various model
parameters with $|\alpha_i| \sim 1$, and in many of the cases found
these pulses to be stable. Thus, we conclude that the stationary
pulse solutions (\ref{NonStaggStationaryPulse}) and
(\ref{StaggStationaryPulse}) to DNLS equation (\ref{3}), (\ref{4})
with parameters satisfying (\ref{stationary}) are generically
stable.

We have also checked the stability of a stationary pulse in the
model where the pulse solution exists only for a selected $\beta$
and, for the pulse placed asymmetrically with respect to the
lattice we found that it is stable. In this simulation, for the
solution (\ref{NonStaggStationaryPulse}) we took the following
pulse parameters $\beta=1/2$, $c=1/4$, and model parameters
$\alpha_2=-0.2553$, $\alpha_4=\alpha_6=-1/2$, and
$\alpha_9=2.2553$ with all other $\alpha_i$ equal to zero. We then
found $\omega = -0.2553$ and $A=0.6557$.

\begin{figure}
\includegraphics{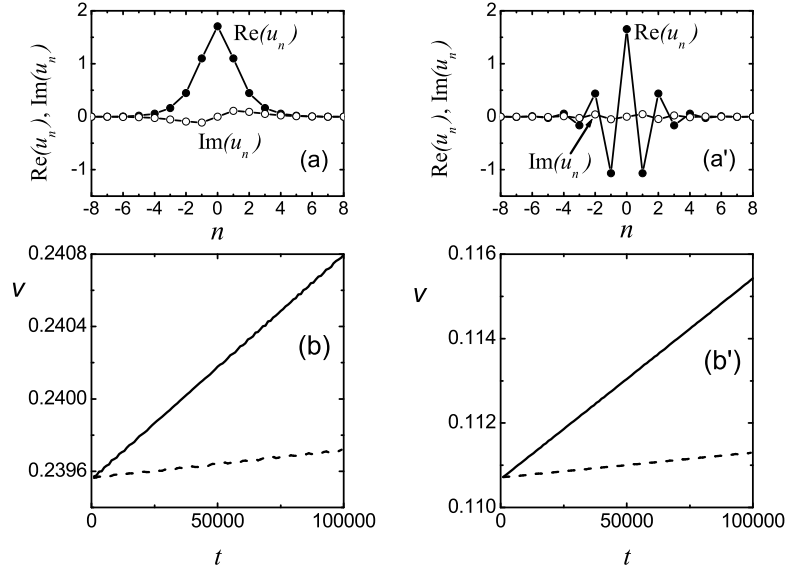}
\caption{(a) Non-staggered moving pulse at $t=0$ and (a') same for
the staggered pulse. In (b) and (b') the long-term evolution of
pulse velocity is shown for the corresponding pulses for the
integration steps of $\tau=5\times 10^{-3}$ (solid line) and
$\tau=2.5\times 10^{-3}$ (dashed line). Numerical scheme with an
accuracy $O(\tau^4)$ is employed. Pulses preserve their velocity
with the accuracy increasing with the increase in the accuracy of
numerical integration. Within the numerical run, the pulse
dynamics is stable in spite of the presence of small perturbations
in the system in the form of rounding errors and integration
scheme errors. The pulse in (a) is given by (\ref{35}),
(\ref{36}), (\ref{41}), and (\ref{38b}). The model and pulse
parameters are as follows: $\alpha_3=-0.473034$, $a_5=1$, and
$\alpha_7=0.946068$; $\beta=1$, $k=0.102102$ (close to 0),
$v=0.239563$, $\omega=-1.07009$, and $A=1.7087$. The pulse in (a')
is given by (\ref{35}), (\ref{36}), (\ref{41}), and (\ref{38c}).
The model and pulse parameters are as follows:
$\alpha_2=0.603116$, $\alpha_3=0.0968843$, and $a_5=0.3$;
$\beta=1$, $k=3.09447$ (close to $\pi$), $v=0.110719$,
$\omega=5.08274$, and $A=1.65172$.} \label{Figure6}
\end{figure}

The robustness of {\em moving} pulse solutions was checked by
observing the evolution of their velocity in a long-term numerical
run. For pulses with amplitudes $A \sim 1$ and velocities $v \sim
0.1$ and for various model parameters supporting the pulse,
$|\alpha_i| \sim 1$, we found that the pulse typically preserves
its velocity with a high accuracy. Two examples of such
simulations, one for the non-staggered pulse and another one for
the staggered pulse are given in Fig. \ref{Figure6} (a), (b) and
(a'), (b'), respectively. In (a) and (a') we show the pulse
configuration at $t=0$ and in (b) and (b') the pulse velocity as
a function of time for two different integration steps,
$\tau=5\times 10^{-3}$ (solid lines) and $\tau=2.5\times 10^{-3}$
(dashed lines), while the numerical scheme with an accuracy
$O(\tau^4)$ is employed.

In both cases, one can notice the linear increase in the pulse
velocity with time, which is due to the numerical error,
since the slope of the line decreases with the decrease in
$\tau$. The presence of perturbation in the form of rounding
errors and integration scheme errors does not result in pulse
instability within the numerical run. Velocity increase rate for
the staggered pulse in (b') is larger than for the non-staggered
one in (b). This can be easily understood because the frequency
of the staggered pulse in almost five times larger than that of
the non-staggered one.

The pulse presented in Fig. \ref{Figure6} (a) is given by
(\ref{38b}). The model has one free parameter and we set
$\alpha_5=1$. For the pulse parameters we set $\beta=1$ and
$k=0.102102$ (close to zero). Then we find from (\ref{35}),
(\ref{36}), (\ref{41}), and (\ref{38b}) the pulse velocity
$v=0.239563$, frequency $\omega=-1.07009$, and amplitude
$A=1.7087$, and the dependent model parameters
$\alpha_3=-0.473034$ and $\alpha_7=0.946068$.

In Fig. \ref{Figure6} (a') the moving pulse solution is given by
(\ref{38c}). The model has one free parameter and we set
$a_5=0.3$. For the pulse parameters we set $\beta=1$ and
$k=3.09447$ (close to $\pi$). Then we find from (\ref{35}),
(\ref{36}), (\ref{41}), and (\ref{38c}) the pulse velocity
$v=0.110719$, frequency $\omega=5.08274$, and amplitude
$A=1.65172$, and the dependent model parameters
$\alpha_2=0.603116$ and $\alpha_3=0.0968843$.

Similar results were observed for the cases when only $\alpha_3$
and $\alpha_5$ are nonzero; only $\alpha_3$ and $\alpha_7$ are
nonzero; only $\alpha_2$, $\alpha_3$, and $\alpha_5$ are nonzero;
only $\alpha_2$, $\alpha_3$, and $\alpha_7$ are nonzero; and  only
$\alpha_2$, $\alpha_5$, and $\alpha_7$ are nonzero.

\begin{figure}
\includegraphics{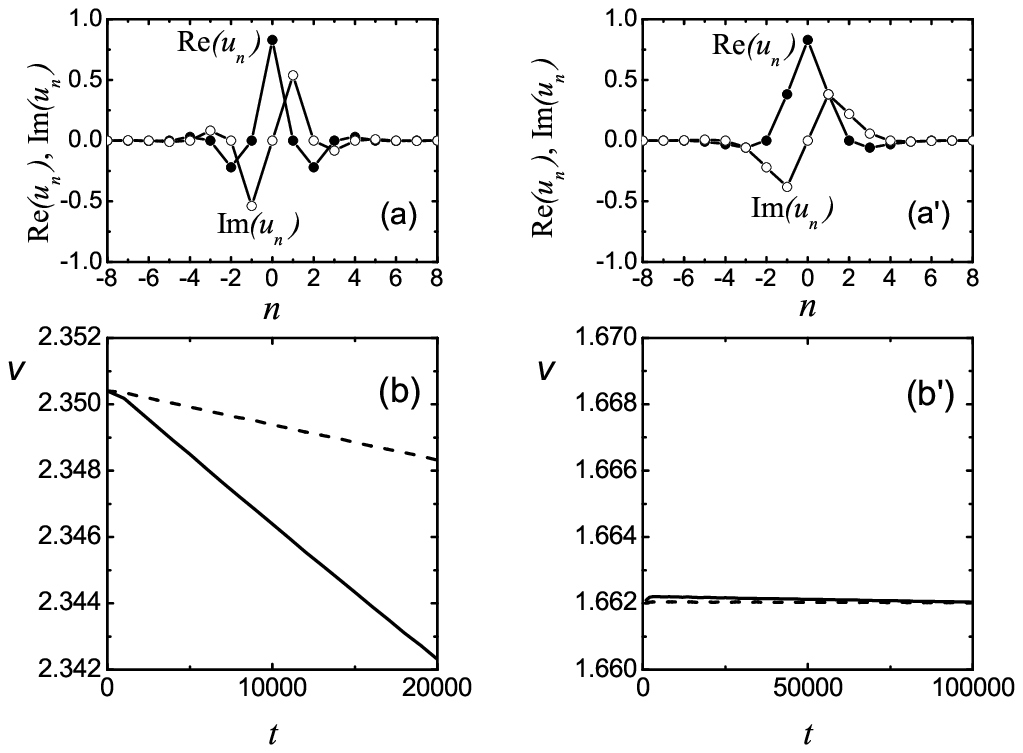}
\caption{Results similar to that shown in Fig. \ref{Figure6} but
for models that are not translationally invariant. (a) and (a')
show the moving pulse profiles at $t=0$. In (b) and (b') the
long-term evolution of pulse velocity is shown for the
corresponding pulses. The integration steps are (b) $\tau=
10^{-3}$ (solid line) and $\tau=5\times 10^{-4}$ (dashed line) and
(b') $\tau=5\times 10^{-3}$ (solid line) and $\tau=2.5\times
10^{-3}$ (dashed line). The pulse in (a) is given by (\ref{35}),
(\ref{36}), (\ref{41}), and (\ref{38d}). The model and pulse
parameters are as follows: $\alpha_2=2$, $a_4=-1/2$, and
$\alpha_6=-1/2$; $\beta=1$, $k=\pi/2$, $v=1.662$,
$\omega=-0.1822$, and $A=0.8310$. The pulse in (a') is given by
(\ref{35}), (\ref{36}), (\ref{41}), and (\ref{38e}). The model and
pulse parameters are as follows: $\alpha_2=2$, $\alpha_9=-1/2$,
and $a_{10}=-1/2$; $\beta=1$, $k=\pi/4$, $v=2.350$, $\omega=2$,
and $A=0.8310$.} \label{Figure7}
\end{figure}

So far we have studied numerically the pulses in the
models with the parameters satisfying
(\ref{stationary}). However, moving pulse solutions exist even in
the case when (\ref{stationary}) is violated. Two such solutions
are presented by (\ref{38d}) and  (\ref{38e}) together with
(\ref{35}), (\ref{36}), and (\ref{41}). As it can be seen from
Fig. \ref{Figure7}, the pulses show a stable long-term dynamics
with pulse velocity being practically constant with the accuracy
increasing with decrease in the step size of numerical
integration. The pulse in (a) is given by (\ref{35}), (\ref{36}),
(\ref{41}), and (\ref{38d}). The model and pulse parameters are as
follows: $\alpha_2=2$, $a_4=-1/2$, and $\alpha_6=-1/2$; $\beta=1$,
$k=\pi/2$, $v=1.662$, $\omega=-0.1822$, and $A=0.8310$. The pulse
in (a') is given by (\ref{35}), (\ref{36}), (\ref{41}), and
(\ref{38e}). The model and pulse parameters are as follows:
$\alpha_2=2$, $\alpha_9=-1/2$, and $a_{10}=-1/2$; $\beta=1$,
$k=\pi/4$, $v=2.350$, $\omega=2$, and $A=0.8310$.

Velocity increase rate in (b) is considerably larger than in (b')
(note the different abscissa scale for these two panels) and this
result can be expected when we take into account that pulse frequency
in (b) is 11 times larger than in (b').

\section{Conclusions and future challenges } \label{Sec:Concl}

For the nine-parameter DNLS equation (\ref{3}), (\ref{4}) with the
continuity constraint (\ref{5}), in Sec. \ref{Sec:dn} and Sec.
\ref{Sec:cn}, we obtained the two moving periodic wave solutions
for the case of $\alpha_1= \alpha_8=0$ (thus, the moving solutions
are supported by the seven-parameter model). The solutions have
the form of ${\rm dn}$ and ${\rm cn}$ Jacobi elliptic functions.
In the limit $m \rightarrow 1$ both solutions reduce to the moving
pulse solution (see Sec. \ref{Sec:Pulse}). We found and described
several sets of model parameters supporting the moving pulse
solution. For the particular choice of model parameters
(\ref{stationary}), the problem of finding stationary solutions is
integrable and the first integral of this problem was given in
Sec. \ref{Sec:TwoPointMap} in the form of a nonlinear map. From
this map {\em any} stationary solution of the corresponding
problem can be constructed iteratively.

We found the stationary pulse solutions to be generically stable,
i.e., for rather arbitrary choice of model parameters $|\alpha_i|
\sim 1$, in many cases, the spectra of the small-amplitude vibrations
calculated for the lattice containing a pulse included no eigenvalues
with positive real parts. In addition, we confirmed the robustness
of moving pulses by observing the pulse velocity evolution in a
long-term numerical run. We found the velocity to be nearly constant
and the deviation from constancy was attributed to the influence of
the accuracy of the numerical integration.  We specifically note that
the moving pulse solutions exist and they exhibit a stable behavior
in long-term numerical runs even for models which do not support
translationally invariant stationary pulse solutions, as demonstrated
in Fig. \ref{Figure7}.

On using the identities for the Jacobi elliptic functions $\cn$ and $\dn$
given below and similar identities for $\sn$, one can similarly obtain
exact solutions of a rather general discrete $\lambda \phi^4$ field
theory with four parameters, as well as of a modified Fermi-Pasta-Ulam
(FPU) model \cite{fpu}, which will be discussed elsewhere.  Our results
are potentially important for optical pulse propagation in glass fibers
and optical waveguides \cite{esm} and time evolution of Bose-Einstein
condensates \cite{tm}.

\newpage

\section{ Appendix} \label{Sec:Appendix}

We list here the various identities for the Jacobi elliptic
functions $\dn(x,m)$ and $\cn(x,m)$ which have been used in
obtaining the various solutions in this paper.

{\bf Identities for $\dn(x,m)$}

 \be\label{14}
 \dn^2(x,m)[\dn(x+a,m)+\dn(x-a,m)]=-\cs^2(a,m)[\dn(x+a,m)+\dn(x-a,m)]
 +2\ns(a,m)\ds(a,m)\dn(x,m)\,,
 \ee
 \be\label{15}
 \dn(x,m)\dn(x+a,m)\dn(x-a,m)=-\cs(a,m)\cs(2a,m)[\dn(x+a,m)+\dn(x-a,m)]
 +\cs^2(a,m)\dn(x,m)\,,
 \ee
 \bea\label{16}
 &&\dn(x,m)[\dn^2(x+a,m)+\dn^2(x-a,m)]=\ds(a,m)\ns(a,m)[\dn(x+a,m)+\dn(x-a,m)]
 \nonumber \\
 &&-2\cs^2(a,m)\dn(x,m)
 +m\cs(a,m)[\cn(x+a,m)\sn(x+a,m)-\cn(x-a,m)\sn(x-a,m)]\,,
 \eea
 \bea\label{17}
 &&\dn(x+a,m)\dn(x-a,m)[\dn(x+a,m)+\dn(x-a,m)] \nonumber \\
 &&=[\ds(2a,m)\ns(2a,m)-\cs^2(2a,m)]
 [\dn(x+a,m)+\dn(x-a,m)] \nonumber \\
 &&+mcs(2a,m)[\cn(x+a,m)\sn(x+a,m)-\cn(x-a,m)\sn(x-a,m)]\,,
 \eea
 \be\label{18}
 \dn^2(x,m)[\dn(x+a,m)-\dn(x-a,m)]=-\cs^2(a,m)[\dn(x+a,m)-\dn(x-a,m)]
 -2m\cs(a,m)\cn(x,m)\sn(x,m)\,,
 \ee
 \bea\label{19}
 &&\dn(x,m)[\dn^2(x+a,m)-\dn^2(x-a,m)]=\ds(a,m)\ns(a,m)[\dn(x+a,m)-\dn(x-a,m)]
 \nonumber \\
 &&+m\cs(a,m)[\cn(x+a,m)\sn(x+a,m)+\cn(x-a,m)\sn(x-a,m)]\,,
 \eea
 \bea\label{20}
 &&\dn(x+a,m)\dn(x-a,m)[\dn(x+a,m)-\dn(x-a,m)] \nonumber \\
&&=[\ds(2a,m)\ns(2a,m)+\cs^2(2a,m)][\dn(x+a,m)-\dn(x-a,m)]
 \nonumber \\
 &&+mcs(2a,m)[\cn(x+a,m)\sn(x+a,m)+\cn(x-a,m)\sn(x-a,m)]\,.
 \eea

{\bf Identities for $cn(x,m)$}

 \be\label{28}
 m\cn^2(x,m)[\cn(x+a,m)+\cn(x-a,m)]=-\ds^2(a,m)[\cn(x+a,m)+\cn(x-a,m)]
 +2\ns(a,m)\cs(a,m)\cn(x,m)\,,
 \ee
 \be\label{29}
 m\cn(x,m)\cn(x+a,m)\cn(x-a,m)=-\ds(a,m)\ds(2a,m)[\cn(x+a,m)+\cn(x-a,m)]
 +\ds^2(a,m)\cn(x,m)\,,
 \ee
 \bea\label{30}
 &&m\cn(x,m)[\cn^2(x+a,m)+\cn^2(x-a,m)]=\cs(a,m)\ns(a,m)[\cn(x+a,m)+\cn(x-a,m)
]
 \nonumber \\
 &&-2\ds^2(a,m)\cn(x,m)
 +ds(a,m)[\dn(x+a,m)\sn(x+a,m)-\dn(x-a,m)\sn(x-a,m)]\,,
 \eea
 \bea\label{31}
 &&m\cn(x+a,m)\cn(x-a,m)[\cn(x+a,m)+\cn(x-a,m)] \nonumber \\
 &&=[\cs(2a,m)\ns(2a,m)-\ds^2(2a,m)]
 [\cn(x+a,m)+\cn(x-a,m)] \nonumber \\
 &&+ds(2a,m)[\dn(x+a,m)\sn(x+a,m)-\dn(x-a,m)\sn(x-a,m)]\,,
 \eea
\be\label{32}
 m\cn^2(x,m)[\cn(x+a,m)-\cn(x-a,m)]=-\ds^2(a,m)[\cn(x+a,m)-\cn(x-a,m)]
 -2\ds(a,m)\dn(x,m)\sn(x,m)\,,
 \ee
 \bea\label{33}
 &&m\cn(x,m)[\cn^2(x+a,m)-\cn^2(x-a,m)]=\cs(a,m)\ns(a,m)[\cn(x+a,m)-\cn(x-a,m)
]
 \nonumber \\
 &&+ds(a,m)[\dn(x+a,m)\sn(x+a,m)+\dn(x-a,m)\sn(x-a,m)]\,,
 \eea
 \bea\label{34}
 &&m\cn(x+a,m)\cn(x-a,m)[\cn(x+a,m)-\cn(x-a,m)] \nonumber \\
&&=[\cs(2a,m)\ns(2a,m)+\ds^2(2a,m)][\cn(x+a,m)-\cn(x-a,m)]
 \nonumber \\
 &&+ds(2a,m)[\dn(x+a,m)\sn(x+a,m)+\dn(x-a,m)\sn(x-a,m)]\,.
 \eea

{\bf Acknowledgment}

A.K. acknowledges the hospitality of the Center for Nonlinear
Studies at LANL.  This work was supported in part by the U.S.
Department of Energy.

\end{document}